\documentclass[aps,prb,twocolumn,superscriptaddress]{revtex4}

\usepackage{graphicx}
\usepackage{amsmath}
\usepackage{amssymb}
\usepackage{braket}

\begin{document}

\title{Efimov Effect in the Dirac Semi-metals}

\author{Pengfei Zhang}
\affiliation{Institute for Advanced Study, Tsinghua University, Beijing, 100084, China}

\author{Hui Zhai}
\affiliation{Institute for Advanced Study, Tsinghua University, Beijing, 100084, China}
\affiliation{Collaborative Innovation Center of Quantum Matter, Beijing, 100084, China}

\date{\today}

\begin{abstract}
Efimov effect refers to quantum states with discrete scaling symmetry and a universal scaling factor, and has attracted considerable interests from nuclear to atomic physics communities. In a Dirac semi-metal, when an electron interacts with a static impurity though a Coulomb interaction, the same scaling of the kinetic and interaction energies also gives rise to such a Efimov effect. However, even when the Fermi energy exactly lies at the Dirac point, the vacuum polarization of electron-hole pair fluctuation can still screen the Coulomb interaction, which leads to derivation from this scaling symmetry and eventually breakdown of the Efimov effect. This distortion of the Efimov bound state energy due to vacuum polarization is a relativistic electron analogy of the Lamb shift for the hydrogen atom. Motivated by recent experimental observations in two- and three-dimensional Dirac semi-metals, in this paper we investigate this many-body correction to the Efimov effect, and answer the question that under what condition a good number of Efimov-like bound states can still be observed in these condensed matter experiments.  

\end{abstract}

\maketitle
\section{Introduction}
The Efimov effect is first proposed by Vitaly Efimov in 1970 for a quantum three-boson system \cite{Efimov}. Solving this three-body problem with a hyper-spherical coordinate, this problem in the vicinity of a two-body resonance can be reduced to a one-dimensional Schr\"odinger equation as
\begin{equation}
\left(-\frac{\hbar^2 d^2}{2m d^2 R}-\frac{\hbar^2}{mR^2}\right)\Psi=E\Psi. \label{non-revl}
\end{equation}
The intriguing feature of Eq. \ref{non-revl} is the presence of a continuous scaling symmetry, that is, by scaling $R\rightarrow \lambda R$ and $E\rightarrow E/\lambda^2$ with an arbitrary number $\lambda$, the Schr\"odinger equation is still satisfied. However, in this case one also has to impose a short-range boundary condition for the wave function to prevent the Thomas collaps, which in general breaks the scaling symmetry. Efimov effect says that all eigenenergies of bound states $E_n$ form a geometric sequence and $E_n \exp(2\pi/s_0)$ is still an eigen-energy, where $s_0$ is a universal constant. This means that the solution with the boundary condition still obeys a discrete scaling symmetry with a universal scaling factor. In the past decade, the Efimov effect has been observed and extensively studied in a number of few-body systems in cold atom systems \cite{cold_atom 1,cold_atom 2,cold_atom 3,cold_atom 4}, as well as the helium trimer \cite{helium}. 

Here we would take the defining property of ``Efimov Effect" as a quantum state or a phenomenon with discrete scaling symmetry and a universal scaling factor, resulting from a system with its Hamiltonian obeying continuous scaling symmetry plus a non-universal boundary condition. Two nontrivial points of this definition are worth highlighting. First, in most cases the boundary condition completely breaks the continuous scaling symmetry, however, in the case of Efimov effect, there remains a discrete scaling symmetry. Second, despite the boundary condition itself is a non-universal one, the scaling factor is still a universal constant that does not depend on the detail of the boundary. With this broad definition, the Efimov effect can be realized in many systems other than few-body systems and can make a broad impact beyond few-body physics, for instance, recently Efimovian expansion has been proposed and observed in dynamics of many-body systems \cite{efimovian_expansion,super efimovian_expansion}.

\begin{figure}[tp]
\includegraphics[width=3.2 in]
{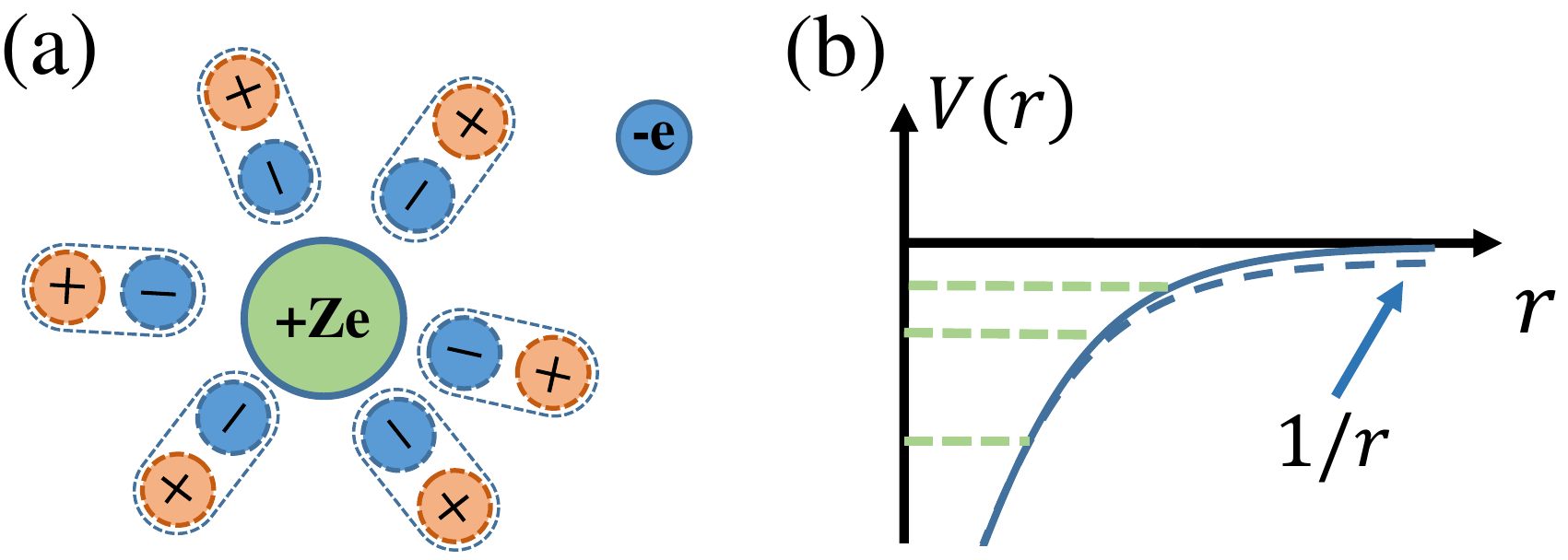}
\caption{(a) Schematic of the electron dipole screening of the interaction potential between an electron and the impurity. (b) Schematic of screened interaction potential with impurity and its effect on Efimov bound state.  }
\label{illustration}
\end{figure}

During recent years, the studies of Dirac and Weyl semi-metals have received considerable attentions in condensed matter physics \cite{review1,review2,review3}. In these systems, the electrons (or holes) have a linear dispersion. Therefore, if one considers a Coulomb interaction between an electron (or a hole) and a static impurity with opposite charge, because the Coulomb interaction and the linear dispersion scale in the same way, the Hamiltonian for this Coulomb impurity problem possesses continuous scaling symmetry, reminiscent of Eq. \ref{non-revl} for the non-relativistic case. Similarly, a short-range boundary condition, depending on the detail of the impurity, breaks the continuous scaling symmetry down to a discrete one. This gives rise to the Efimov effect \cite{Efimov_Dirac Nishida1,Efimov_Dirac Nishida2,Efimov_Dirac 07 2,Efimov_Dirac 07 3}. A recent experiment reported a log-periodic magnetoresistenance oscillation in a potential three-dimensional Dirac semi-metal ZrTe$_5$ \cite{ExpPKU}. They attribute this observation to such Efimov effect in the Dirac semi-metals \cite{ExpPKU}. Similar experimental evidence from local tunneling measurement nearby charged impurities has also been reported in grephene as a two-dimensional Dirac semi-metal \cite{ExpIsrael}. To the best of our knowledge, these two are first experimental manifestations of the Efimov effect in condensed matter systems.

However, in the consideration so far, the Coulomb interaction between electrons has been ignored. This Coulomb interaction  can screen the interaction between electron and impurity, and consequently distorts or even destroys the Efimov effect. In this paper we address the question that to what extent the Efimov effect can survive after including the many-body interactions. Given the fact that such Coulomb interaction is certainly present in current experiments \cite{ExpPKU,ExpIsrael}, this question becomes of crucial importance for explaining these observations. 

To lay the basis, we start with a simplified situation with a single Dirac point and the chemical potential right at the Dirac point, and the electrons interact with a static impurity with charge $Ze$. Though charge neutral, the electronic dipoles made of electron-hole pairs, i.e. the vacuum polarization, can still screen the interaction potential with impurity, as schematized   in Fig. \ref{illustration}(a), and the behavior of the screened potential will no longer be proportional to $-1/r$, as schematically shown in Fig. \ref{illustration}(b) \cite{Efimov_Dirac 07 1,Nagaosa,Yao}. Thus the screened potential losses the scale invariance, and moreover, because the shallow Efimov bound state is quite sensitive to the long-range part of the potential, these shallow bound state will first disappear as the screening effect is turned on. 

In some sense, it can be viewed as a close analogy of the Lamb shift, where the vacuum polarization modifies the Coulomb potential between electron and nucleus, and therefore distorts the energy level of a hydrogen atom. But there is a significant difference. Physically, in the case of Lamb shift, the electrons have a finite mass and can be approximated as non-relativistic but in this case the electrons are gapless and is essentially relativistic. Apparently, for the relativistic case the energy level is more sensitive to this modification of the Coulomb potential. Technically, the traditional Lamb shift consider the shift of the energy of a bound state and directly compute the self-energy from radiation while here we are considering a quasi-bound state and we choose to use a Wilson renormalization scheme which selects certain diagrams.

\begin{figure}[tp]
\includegraphics[width=3.2 in]
{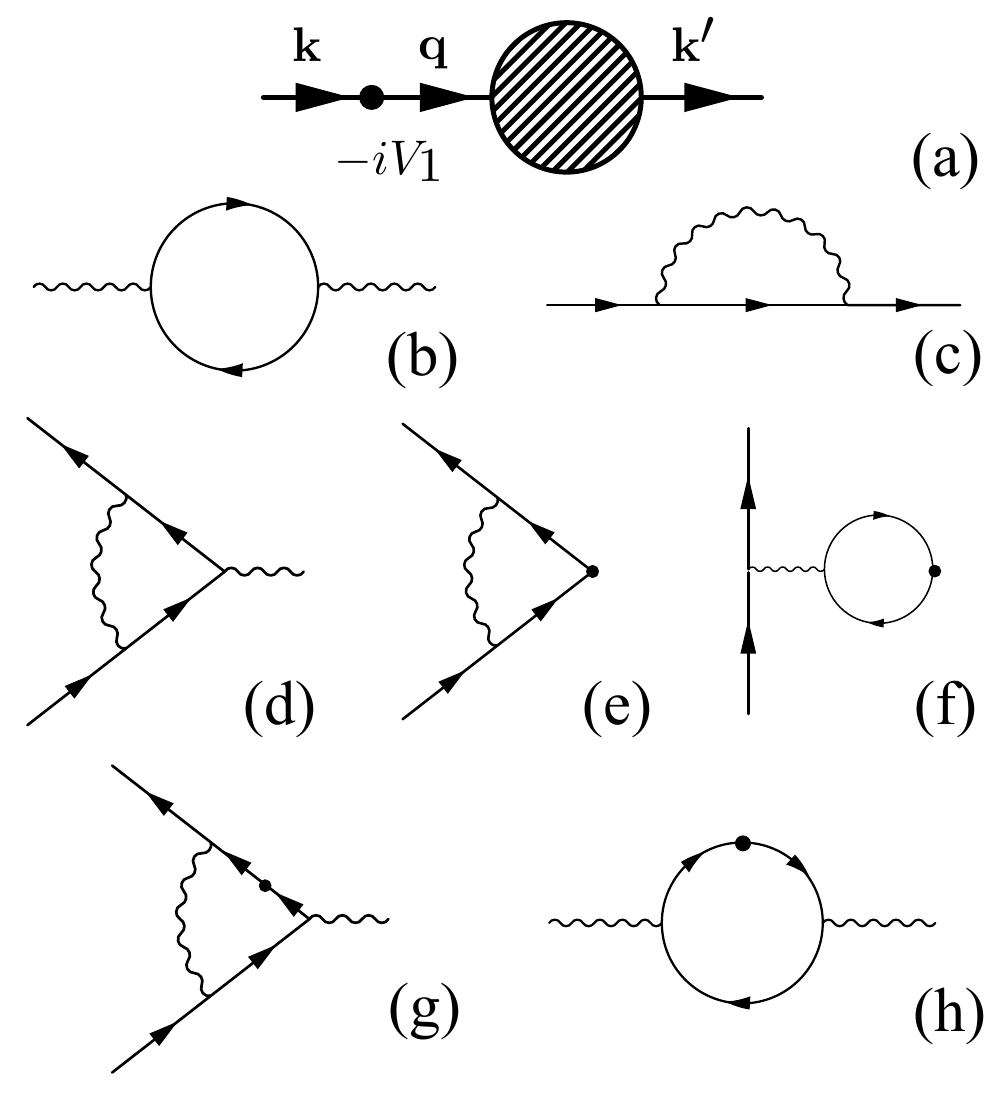}
\caption{(a) Diagram for scattering amplitude between electron and impurity. (b-c) Self-energy digrams for photon (b) and for electron (c). (d-e) One-loop correction for single photon emission (d) and for single particle scattering vertex (e and f). (g) Diagram for single photon scattering. (h) Diagram for single photon emission with non-zero total momentum.}
\label{diagram}
\end{figure}

\section{Field Theory Model} 
To quantitatively address this issue, we employ a field theory approach. This is because, on one hand, the Efimov effect can be captured by limit cycle solutions of the renormalization group (RG); and on the other hand, it is also a natural way to include the screening effect due to vacuum polarization. Thus, it provides a natural framework to combine both effects. By fixing $v_\text{F}=1$, the Lagrangian for a Dirac semi-metal at $d=3$ and $d=2$ can be written as
\begin{align}
\mathcal{L}=\int dt d^{d}{\bf r}\left(\bar{\psi}(i\partial_\mu\gamma^\mu)\psi-\bar{\psi}\gamma^0\psi V({\bf r})-e\bar{\psi}\gamma^0\psi \phi\right)+\mathcal{L}_\phi,
\end{align} 
where $\mu=0,1,\dots,d$. For $d=3$, $\psi$ is a four component fermion field and 
\begin{eqnarray}
\gamma^\mu=\begin{pmatrix}
0&\sigma^\mu\\
\bar{\sigma}^\mu&0
\end{pmatrix},
\ \ \ \ \sigma^\mu=(I,\mathbf{\sigma}),\ \ \ \ \bar{\sigma}^\mu=(I,-\mathbf{\sigma}).
\end{eqnarray}
For $d=2$, $\psi$ is a two-component fermion field and $\gamma^0=\sigma_z$, $\gamma^1=i\sigma_y$ and $\gamma^2=-i\sigma_x$. Thus, the first term gives rise to a linear dispersion in $d$-dimension. In the second term, $V({\bf r})=-V_\text{c}/|{\bf r}|$ is the Coulomb interaction between electron and the static impurity at ${\bf r}=0$ and $V_\text{c}=Z\alpha$, $e^2=4\pi\alpha$ at the bare level. $\phi$ describes the photon field mediating the instantaneous Coulomb interaction. The third term describes the coupling between fermion and photon. The last term represents the Lagrangian for the photon field. For $d=3$,
\begin{equation}
\mathcal{L}_\phi=\int dt d^3{\bf r}\frac{1}{2}(\nabla \phi)^2.
\end{equation}
And for $d=2$, knowing the Fourier transformation
 \begin{align}
 \int d^2r\frac{1}{r}\exp(i\mathbf{q}\cdot\mathbf{r})=\frac{2\pi}{|\mathbf{q}|},
 \end{align}
we have
\begin{align}
\mathcal{L}_\phi=\int dt d^2{\bf q}|{\bf q}|\phi({\bf q})\phi(-{\bf q}).
\end{align}

When Efimov effect occurs, the wave function and the eigen-energy is quite sensitive to the short range details. To include this effect, we should add an extra term in the effective field theory
\begin{equation}
\mathcal{L}^\prime=-\int dtd^d{\bf r} V_\text{s} \bar{\psi}\gamma^0\psi\delta^d({\bf r}).
\end{equation}
$V_\text{s}$ denotes a strength of this short-range interaction. In the RG analysis below, $V_\text{s}$ flows as a function of the energy cut-off $\Lambda$. The energy scale $\Lambda$ at which $V_\text{s}$ diverges corresponds to the eigen-energy of a bound state \cite{Braaten}. A limit cycle behavior of this RG flows manifests the Efimov effect. 

\section{RG Flow without Screening Effect} 
\begin{figure}[tbp]
	\includegraphics[width=0.45\textwidth]{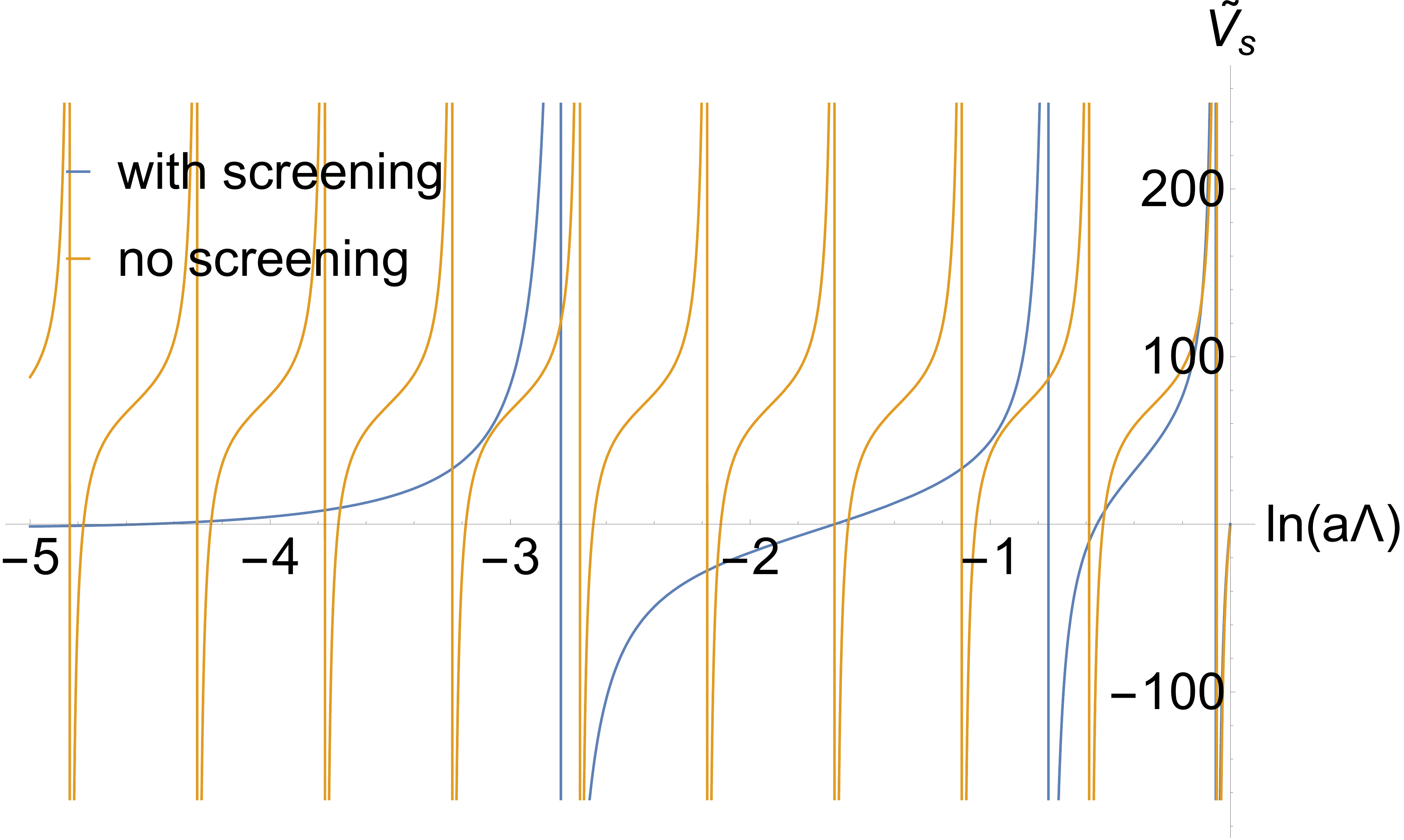}
	\caption{A typical flow diagram for $\tilde{V}_\text{s}$ for the case without screening effect (yellow dashed line) and for the case with screening effect (blue solid line). Here $d=3$, $Z=2$ and $\alpha=3$ }
	\label{flow}
\end{figure}

Before discussion the screening effect, let us first turn off the coupling between electron and photon, and review how to formulate the Efimov effect in this Coulomb impurity in term of RG flows as follows:

As shown in Fig. \ref{diagram}(a), defining the scattering amplitude between an electron and the impurity $M(\mathbf{k'},\mathbf{k},E)$ where $E$ is the energy, $\bf{k}$ is the incoming momenta and $\bf{k}^\prime$ is the outgoing momenta,  $M(\mathbf{k'},\mathbf{k},E)$ satisfies a self-consistent equation that reads
\begin{align}
&iM(\mathbf{k'},\mathbf{k},E)=-iV_1(\mathbf{k'},\mathbf{k})\gamma^0+\notag\\&\int_0^{\Lambda}\frac{d^3q}{(2\pi)^3}(-i)V_1(\mathbf{q},\mathbf{k})iM(\mathbf{k'},\mathbf{q},E)\frac{iq_\mu\gamma^\mu}{E^2-q^2+i\epsilon} \gamma^0, \label{M}
\end{align}
where we have defined the one-particle irreducible scattering vertex
\begin{equation}
V_1(\mathbf{k'},\mathbf{k})=\left(V_\text{s}-\frac{4\pi V_\text{c}}{(\mathbf{k}-\mathbf{k'})^2}\right).
\end{equation}
Since the scattering amplitude is a physical observable, it should not depend on the high energy cutoff $\Lambda$. Thus, we should choose $V_\text{s}$ in such a way that Eq. \ref{M} results in a $\Lambda$-independent $M(\mathbf{k'},\mathbf{k},E)$. With this requirement and after some lengthy derivation \cite{supp}, it final gives that for $d=3$
\begin{align}
\tilde{V}_\text{s}=-4\pi V_\text{c}\frac{\cos(s_0\ln(\Lambda/\Lambda_0)+\phi)}{\cos(s_0\ln(\Lambda/\Lambda_0)-\phi)},\label{Vsd3}
\end{align}
where $\tilde{V}_\text{s}=V_\text{s}\Lambda^2$, $s_0=\sqrt{V_\text{c}^2-1}$, $\tan\phi=s_0$ and $\Lambda_0$ is a fixed energy scale responsible for the non-universal short-range physics.
And for $d=2$, it gives
\begin{align}
\tilde{V}_\text{s}=-2\pi V_\text{c}\frac{\cos(s_0\ln(\Lambda/\Lambda_0)+\varphi)}{\cos(s_0\ln(\Lambda/\Lambda_0)-\varphi)}, \label{Vsd2}
\end{align}
where $\tilde{V}_\text{s}=V_\text{s}\Lambda$, $s_0=\sqrt{V_0^2-1/4}$ and $\tan\varphi=2s_0$.
In both cases, $\tilde{V}_\text{s}$ should log-periodic structure, as shown in Fig. \ref{flow}, and by scaling $\Lambda\rightarrow e^{-n\pi/s_0}\Lambda$, $\tilde{V}_\text{s}$ is invariant. 

The renormalization relations Eq. \ref{Vsd3} and Eq. \ref{Vsd2} can be used to derive an exact RG flow equation by requiring no explicit $\Lambda$-dependence on the r.h.s. of the RG equation, that gives 
\begin{align}
\frac{d\tilde{V}_\text{s}}{d\ln \Lambda}=8\pi\frac{V_\text{c}^2-1}{V_\text{c}}\left(1+\frac{(4\pi(2-V_\text{c}^2)+\tilde{V}_\text{s}V_\text{c})^2}{64\pi^2(V_\text{c}^2-1)}\right) \label{RGexactd3}
\end{align}
for $d=3$ and 
\begin{align}
\frac{d\tilde{V}_\text{s}}{d\ln \Lambda}=&2\pi s_0 V_\text{c} \sin(2\varphi)\times\nonumber\\
&\left(\frac{(2\pi V_\text{c}(1-4 s_0^2)+\tilde{V}_\text{s}(1+4s_0^2))^2}{64\pi^2s_0^2V_\text{c}^2}+1\right)
\label{RGexactd2}
\end{align}
for $d=2$. It is easy to show that the solution of Eq. \ref{RGexactd3} and Eq. \ref{RGexactd2} reproduce Eq. \ref{Vsd3} and Eq. \ref{Vsd2}, respectively.

\section{RG Flow with Screening Effect}
 Now we turn on the coupling between electron and photon. The stratagem here is that the interaction between electron and impurity is treated exactly as discussed above, while the interaction between electron and photon is treated perturbatively. Since the RG equation should recover the exact one when electron-photon coupling is turned off, we should therefore add diagram from the perturbative calculation onto the exact RG equation derived above. 

The principles that we select diagrams for electron-photon interaction are listed as follows: (i) Since in the Wilson renormalization group calculation, at each step we only perform momentum integration over a thin shell with width $d\Lambda$ at energy $\Lambda$, thus, to the order of $O(d\Lambda/\Lambda)$, we only consider the one-loop diagrams for irreducible self-energy or vertex which only contains a single momentum integral. This selects out diagrams shown in Fig. \ref{diagram} (b)-(g). (ii) We only keep diagrams that are relevant or marginal under power counting. These diagrams give leading contribution when we go the the low-energy limit. As a result, the diagram (g) describing the emission of a photon with non-zero total momentum can be neglected, because to the one-loop order and for small total momentum, it gives rise to 
\begin{align}
\int dt d\mathbf{p_1}d\mathbf{p_2}d\mathbf{p_3} \phi(\mathbf{p_1})\bar{\psi}(\mathbf{p_2})\gamma^0\psi(\mathbf{p_3})\frac{e'}{(\mathbf{p_3+p_1-p_2})^2},
\end{align}
which is irrelevant. (iii) The diagram (h) represents the single-body scattering of photon, and it is indeed zero since it is equal to a three-photon scattering diagram which vanishes with the presence of particle-hole symmetry. Hence, we only need to include the contribution (b-f) in the Fig. \ref{diagram}. 

With these diagrams and following the standard Wilson RG procedure \cite{supp}, one arrives at the following RG flow equations for $d=3$ case
\begin{align}
&\frac{d\tilde{V}_\text{s}}{d\ln \Lambda}=\text{(R.H.S. of Eq. \ref{RGexactd3})}+\frac{V_\text{c}e^2}{6\pi}+\frac{2\tilde{V}_\text{s}e^2}{3\pi^2}, \label{perRGd3}
\\ &\frac{dV_\text{c}}{d\ln \Lambda}=\frac{V_\text{c}e^2}{3\pi^2},\label{perRGd3V}\\ 
&\frac{de}{d\ln \Lambda}=\frac{e^3}{6\pi^2}.\label{perRGd3e}
\end{align}
Here we choose initial condition for the RG flow starting at the momentum scale $\Lambda\sim1/a$, where $a$ is treated as the lattice energy length, at which the bare value $V_\text{c}(1/a)=Z\alpha$, $\tilde{V}_\text{s}(1/a)=0$, $e(1/a)=\sqrt{4\pi\alpha}\equiv e_0$ and $\tilde{V}_\text{s}(1/a)=0$. One can solve the last two equations analytically, which give:
\begin{align}
e^2(\Lambda)&=\frac{e_0^2}{1-e_0^2\ln(\Lambda a)/3\pi^2}\\
V_c(\Lambda)&=\frac{Z\alpha}{1-e_0^2\ln(\Lambda a)/3\pi^2}
\end{align} 
This comes from two effects. First although zero density of states near the Fermi surface leads to an infinite screening length, there should still be a $\ln r$ screening for the Coulomb potential by the particle-hole excitations. Second, the Fermi velocity actually flows under the RG, however, since we choose to fix $v_F=1$, this make $V_0$ smaller. 

\begin{figure}[t]
	\includegraphics[width=0.4\textwidth]{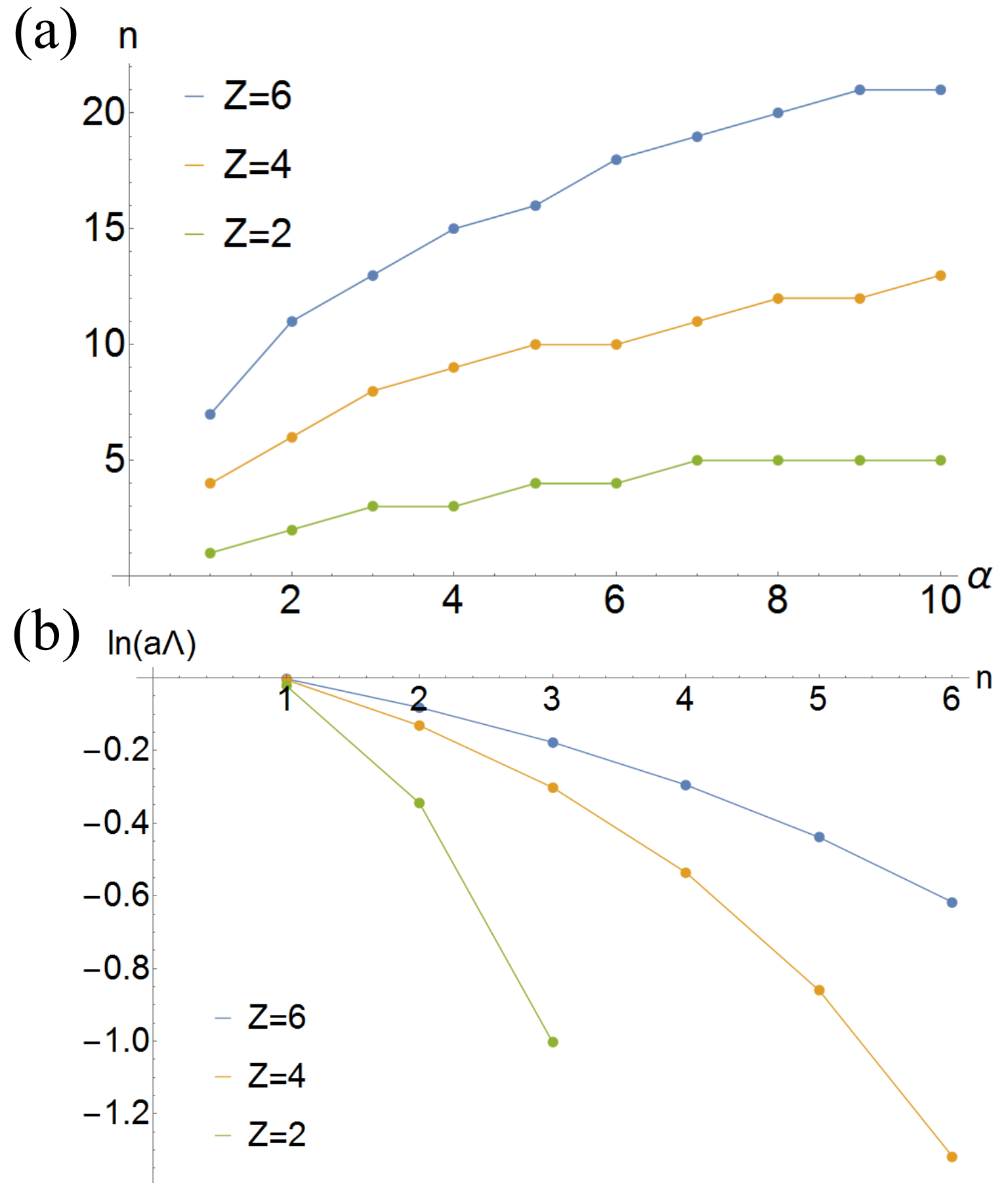}
	\caption{Dirac semi-metal at $d=3$ (a): The number of bound state v.s. $\alpha$ for three different values of $Z$. (b) The bound state energy $E_n$, plotted in a dimensionless form $\ln (aE_n)$, for different $n$ and with three different values of $Z$. $\alpha$ is fixed at $\alpha=5$. }
	\label{d_3result}
\end{figure}

As we show in Fig. \ref{flow}, because of the presence of the extra terms in Eq. \ref{perRGd3}, the limit cycle solution is destroyed after a few periods. Moreover the decrease of $V_0$ as lowering $\Lambda$ gives a larger periodicity since $s_0(\Lambda)\sim\sqrt{V_0^2(\Lambda)-1}$. The energies of the bound state no longer obey a perfect geometric sequence. 

Similarly, for $d=2$, we can obtain 
\begin{align}
\frac{d\tilde{V}_\text{s}}{d\ln \Lambda}=\text{(R.H.S. of Eq. \ref{RGexactd2})}+\frac{e^2 \tilde{V}_\text{s}}{8\pi}, \label{perRGd2}
\end{align}
with
\begin{align}
e^2(\Lambda)=\frac{e_0^2}{1-e_0^2 \ln \Lambda a/ 16\pi},\\
V_\text{c}(\Lambda)=\frac{Z\alpha}{1-e_0^2 \ln \Lambda a/ 16\pi},
\end{align}
where the only correction comes from diagram (c) in Fig. \ref{diagram} which renormalizes the Fermi velocity. All the other diagrams give a correction of the order of ${\bf q}^2$ at low energy which is smaller comparing to $|\bf{q}|$ behavior of the photon propagator and the impurity vertex, or the constant coupling of single photon emission.

\begin{figure}[tbp]
	\includegraphics[width=0.4\textwidth]{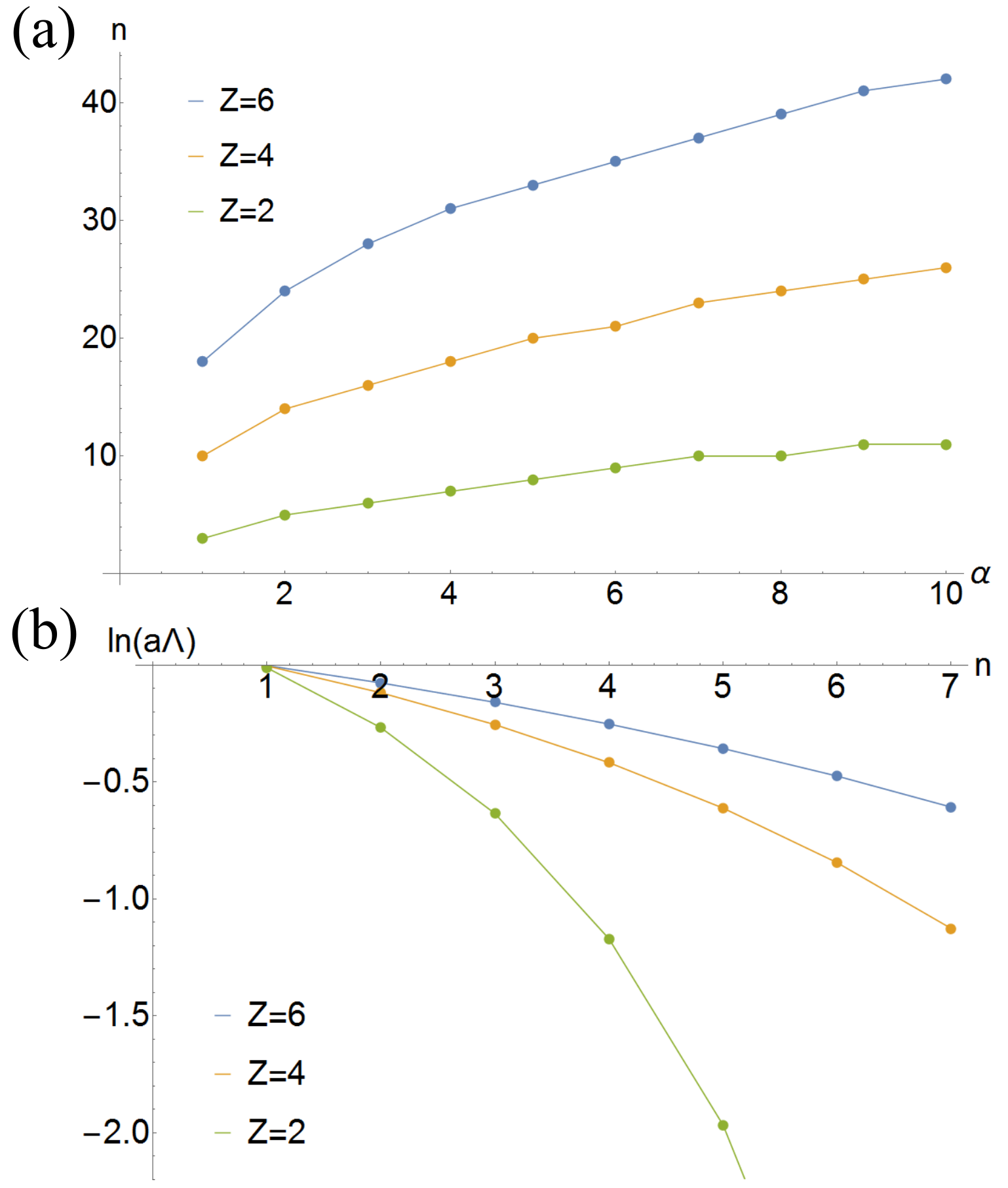}
	\caption{The same plot as Fig. \ref{d_3result} but for $d=2$ Dirac semi-metal. Here $\alpha$ is also fixed at $\alpha=5$. }
	\label{d_2result}
\end{figure}

The final results of this work are presented in Fig. \ref{d_3result} and Fig. \ref{d_2result} for $d=3$ and $d=2$ Dirac seme-metals, respectively. Essentially $Z$ and $\alpha$ are only two parameters in these systems. In Fig. \ref{d_3result}(a) and Fig. \ref{d_2result}(a), we show the number of bound states for various $Z$ and $\alpha$, and in Fig. \ref{d_3result}(b) and Fig. \ref{d_2result}(b), we show to what extent the bound state energies obey a geometric sequence. It is very clear that in both cases, the larger $Z$ the more stable Efimov effect. This can be roughly understood as that the correction to $V_\text{c}$ is controlled by $e_0^2\ln(E_n a)\sim e_0^2/s_0\sim1/Z$, and this correction is smaller for a larger $Z$. In the experiment of Ref. \cite{ExpPKU} from a three-dimensional Dirac semi-metal, about five logarithmic period have been observed in the magnetoresistenance oscillation. Our results suggest that a larger $Z$ is needed in order to reach such a regime and the experimental observation is more likely due to a static impurity with a large charge in the material. 

\section{summary}

In conclusion, we have studied the many-body correction of vacuum polarization to the Efimov effect between a Dirac electron and a charged impurity, which is an analogy of the Lamb shift for ``relativistic hydrogen atoms". Our results are directly related to recent transport experiments on Dirac materials, and can also be applied to Weyl materials. This paves a way toward more solid understanding of the emergent Efimov effect in condensed matter systems. 

\textit{Acknowledgment.} We would like to thank Hong Yao, Ran Qi, Jian Wang and Haiwen Liu for helpful discussion. This work is supported by MOST under Grant No. 2016YFA0301600 and NSFC Grant No. 11325418 and No. 11734010.

\appendix 
\begin{widetext}

\section{Derivation of perturbative RG equations}\label{A}
In this appendix, we outline the calculation that lead to the renormalization relations \eqref{perRGd3}, \eqref{perRGd3V} and \eqref{perRGd3e} for Dirac fermions in $d=3$. The calculation in $d=2$ is similar. We choose to employ a perturbative Wilson RG scheme with a cutoff $\Lambda$ in momentum space. In other words, the integration over momentum $\mathbf{q}$ is performed over a momentum shell $\Lambda-d\Lambda<|\mathbf{q}|<\Lambda$ while the frequency is integrated from $-\infty$ to $\infty$. The diagrams we considered are shown in Fig. \ref{diagram} as explained in the main text. 

Before integrating out the high-energy fluctuation, we have an action at cut-off $\Lambda$:
\begin{align}
S^{(\Lambda)}=\int d^3x dt \left(\bar{\psi}(i\partial_\mu\gamma^\mu)\psi+\bar{\psi}\gamma^0\psi V_c(\Lambda)/|\mathbf{r}|-e(\Lambda)\bar{\psi}\gamma^0\psi \phi+\frac{1}{2}(\nabla\phi)^2-V_s(\Lambda)\bar{\psi}\gamma^0\psi\delta^{(3)}(\mathbf{x})\right).
\end{align} 
 After integrating out the virtual particles in the moementum shell and performing a rescaling, we could derive the effective action defined with cutoff $\Lambda-d\Lambda$. Then we find the RG equations for $\tilde{V}_s$, $V_c$ and $e$ by comparing it with the definition of $S^{(\Lambda-d\Lambda)}$. 

 Firstly we need to include the renormalization of photon propagator. This is given by the self energy $\Pi(\omega,\mathbf{p})$ of photons as shown in the diagram in Fig. \ref{diagram} (b). Using the Feynman rules, we have
\begin{align}
i\Pi(\omega,\mathbf{p})=&-\int\frac{d^4q}{(2\pi)^4}e^2\frac{\text{tr}\left[\gamma^0\gamma^\mu\gamma^0\gamma^\nu\right]}{(\omega+q_0)^2-(\mathbf{p+q})^2+i\epsilon}\frac{(p+q)_{\mu}q_\nu}{(q_0)^2-(\mathbf{q})^2+i\epsilon}\notag\\=&-\int\frac{d^4q}{(2\pi)^4}\frac{(\omega+q_0)q_0+(\mathbf{p+q})\cdot\mathbf{q}}{(\omega+q_0)^2-(\mathbf{p+q})^2+i\epsilon}\frac{4e^2}{(q_0)^2-(\mathbf{q})^2+i\epsilon}.
\end{align}
We keep the leading order contribution for small $\omega$ and $\mathbf{p}$ by the Taylor expansion. Firstly we set $\mathbf{p}=0$ and this gives the result
\begin{align}
i\Pi(\omega,\mathbf{0})\propto\int d^3q\left(\frac{(-q(-q-\omega)+q^2)}{((q+\omega)^2-q^2)(-2q)}+\frac{(-q(-q+\omega)+q^2)}{((-q+\omega)^2-q^2)(-2q)}\right)=0.
\end{align}
after integrating out $q_0$ using contour integral. This means we only have terms proportional to $\mathbf{p}^2$, which gives the field strength renormalization for $\phi$. Thus we set $\omega=0$, and expand $\Pi(0,\mathbf{p})$ to the order of $\mathbf{p}^2$:
\begin{align}
i\Pi(0,\mathbf{p})=-\int\frac{d^4q}{(2\pi)^4}\frac{q_0^2+(\mathbf{p+q})\cdot\mathbf{q}}{q_0^2-(\mathbf{p+q})^2+i\epsilon}\frac{4e^2}{q_0^2-(\mathbf{q})^2+i\epsilon}=i\frac{e^2}{6\pi^2}\mathbf{p}^2d\ln\Lambda.
\end{align}

The contribution to the renormalization of the single-particle scattering vertex shown diagram in Fig. \ref{diagram} (f) indeed gives the same result. Another contribution to the single-particle scattering is given by diagram (e), this gives 
\begin{align}
iM_s(0,\mathbf{k})(i\frac{4\pi V_c}{k^2}-iV_s)=\int\frac{d^4q}{(2\pi)^4}(-ie\gamma^0)\frac{i(q+k)_\mu\gamma^\mu}{(q+k)_\mu(q+k)^\mu}\gamma^0\frac{iq_\mu\gamma^\mu}{q_\mu q^\mu}(-ie\gamma^0)(i\frac{4\pi V_c}{k^2}-iV_s)\frac{i}{\mathbf{q}^2}.
\end{align}
Here we define $k_\mu=(0,\mathbf{k})$ and $k=|\mathbf{k}|$. After expanding to the lowest order of $k$, we find the result is:
\begin{align}
(i\frac{4\pi V_c}{k^2}-iV_s)\frac{e^2}{24\pi^2}.
\end{align}

The renormalization of the photon emission vertex is shown in Fig. \ref{diagram} (d). Expanding to the zero-th order of frequency and momentum , it is just given by $iM_s(0,\mathbf{0})=0$. 

At last we consider the self energy $D(\omega,\mathbf{k})$ shown in Fig. \ref{diagram} (c). This diagram renormalizes propagator for the fermion as $i/(k_\mu\gamma^\mu-D)$:
\begin{align}
-iD(\omega,\mathbf{k})=\int\frac{d^4q}{(2\pi)^4}\frac{e^2}{\mathbf{q}^2}\frac{q_0 I-(q+k)^i\gamma^i}{q_0^2-(\mathbf{q+k})^2+i\epsilon}.
\end{align}
There is no $\omega$ dependence, and the constant part is indeed non-physical which should be canceled by tuning the chemical potential. We only consider the momentum dependent part and find
\begin{align}
-i(D(\omega,\mathbf{k})-D(\omega,\mathbf{0}))=k_i \gamma^i\frac{ie^2}{6\pi^2}d\ln\Lambda.
\end{align}

Adding all these contributions together, we find the effective action after integrating out virtual particles:
\begin{align}
S^{(\Lambda-d\Lambda)}_{\text{eff}}=\int d^3x dt &\bar{\psi}(i\partial_0\gamma^0+i\partial_i\gamma^i(1+\frac{e^2}{6\pi^2}d\ln \Lambda))\psi+\bar{\psi}\gamma^0\psi \frac{V_c(\Lambda)}{(1+\frac{e^2}{6\pi^2}d\ln \Lambda)|\mathbf{x}|}\notag\\&-e(\Lambda)\bar{\psi}\gamma^0\psi \phi+\frac{1}{2}(1+\frac{e^2}{6\pi^2}d\ln \Lambda)(\nabla\phi)^2-V_s(\Lambda)\frac{(1-\frac{V_ce^2}{6\pi}d\ln\Lambda)}{(1+\frac{e^2}{6\pi^2}d\ln \Lambda))}\bar{\psi}\gamma^0\psi\delta^{(3)}(\mathbf{x}).
\end{align} 
To bring it back to the form of $S^{(\Lambda-d\Lambda)}$, we rescale $x\rightarrow(1+\frac{e^2}{6\pi^2}d\ln \Lambda)x$, $\psi\rightarrow(1+\frac{e^2}{6\pi^2}d\ln \Lambda)^{-3/2}\psi$ and $\phi\rightarrow(1+\frac{e^2}{6\pi^2}d\ln \Lambda)^{-1}\phi$. Finally, we find the renormalization relations:
\begin{align}
\frac{d\tilde{V}_s}{d\ln \Lambda}&=8\pi\frac{V_c^2-1}{V_c}\left(1+\frac{(4\pi(2-V_c^2)+\tilde{V}_sV_c)^2}{64\pi^2(V_c^2-1)}\right)+\frac{V_ce^2}{6\pi}+\frac{2\tilde{V}_se^2}{3\pi^2}, \label{perRG}
\\\frac{dV_c}{d\ln \Lambda}&=\frac{V_ce^2}{3\pi^2},\ \ \ \ \ \frac{de}{d\ln \Lambda}=\frac{e^3}{6\pi^2}.
\end{align}
\end{widetext}


\begin{thebibliography}{99}

\bibitem{Efimov}
V. Efimov. Phys. Lett. 33B, \textbf{563} (1970).

\bibitem{cold_atom 1}
T. Kraemer, M. Mark, P. Waldburger, J. G. Danzl, C. Chin, B. Engeser, A. D. Lange, K. Pilch, A. Jaakkola, H.-C. Nägerl and R. Grimm, Nature \textbf{440}, 315-318 (2006)

\bibitem{cold_atom 2}
B. Huang, L. A. Sidorenkov, R. Grimm, J. M. Hutson, Phys. Rev. Lett. \textbf{112}, 190401 (2014).

\bibitem{cold_atom 3}
R. Pires, J. Ulmanis, S. Häfner, M. Repp, A. Arias, E. D. Kuhnle, M. Weidemüller, Phys. Rev. Lett. \textbf{112}, 250404 (2014).

\bibitem{cold_atom 4}
S.-K. Tung, K. Jiménez-García, J. Johansen, C. V. Parker, C. Chin, Phys. Rev. Lett. \textbf{113}, 240402 (2014).

\bibitem{helium}
M. Kunitski, S. Zeller, J. Voigtsberger, A. Kalinin, L. Ph. H. Schmidt, M. Schöffler, A. Czasch, W. Schöllkopf, R. E. Grisenti, T. Jahnke, D. Blume and R. Dörner, Science \textbf{348}, 6234, 551-555 (2015).

\bibitem{efimovian_expansion}
S. Deng, Z.-Y. Shi, P. Diao, Q. Yu, H. Zhai, R. Qi, H. Wu, Science \textbf{353}, 6297, 371-374 (2016).

\bibitem{super efimovian_expansion}
Z.-Y. Shi, R. Qi, H. Zhai, Z. Yu, arXiv:1608.05799.

\bibitem{review1}
A. H. Castro Neto, F. Guinea, N. M. R. Peres, K. S. Novoselov, and A. K. Geim, Rev. Mod. Phys. \textbf{81}, 109 (2009).

\bibitem{review2}
M. Z. Hasan and C. L. Kane, Rev. Mod. Phys. \textbf{82}, 3045 (2010).

\bibitem{review3}
 X. L. Qi and S. C. Zhang, Rev. Mod. Phys. \textbf{83}, 1057 (2011).

\bibitem{Efimov_Dirac 07 2}
A. V. Shytov, M. I. Katsnelson and L. S. Levitov, Phys. Rev. Lett. \textbf{99}, 246802 (2007).

\bibitem{Efimov_Dirac 07 3}
V. M. Pereira, J. Nilsson, and A. H. C. Neto, Phys. Rev. Lett. \textbf{99}, 166802 (2007).

\bibitem{Efimov_Dirac Nishida1}
Y. Nishida, Phys. Rev. B \textbf{90}, 165414 (2014).

\bibitem{Efimov_Dirac Nishida2}
Y. Nishida, Phys. Rev. B \textbf{94}, 085430 (2016).

\bibitem{ExpPKU}
H. Wang, H. Liu, Y. Li, Y. Liu, J. Wang, J. Liu, Y. Wang, L. Li, J. Yan, D. Mandrus, X. C. Xie, J. Wang, \emph{arXiv}:1704.00995.

\bibitem{ExpIsrael}
O. Ovdat, J. Mao, Y. Jiang, E. Y. Andrei, E. Akkermans, Nature Communications \textbf{8}, 507 (2017).

\bibitem{Efimov_Dirac 07 1}
A. V. Shytov, M. I. Katsnelson, and L. S. Levitov, Phys. Rev. Lett. \textbf{99}, 236801 (2007).

\bibitem{Nagaosa}
H. Isobe and N. Nagaosa, Phys. Rev. B \textbf{86}, 165127 (2012).

\bibitem{Yao}
S.-K. Jian and H. Yao, Phys. Rev. B \textbf{92}, 045121 (2015).

\bibitem{Braaten}
E. Braaten and H.-W. Hammer, Physics Reports, \textbf{428}: 259-390 (2006).

\bibitem{supp}
See Appendix. \ref{A} for detailed derivation. 



\end{thebibliography}
\end{document}